\documentclass[12pt]{article}
\pdfoutput=1 
\usepackage{amsmath}
\usepackage[table,xcdraw]{xcolor}
\usepackage[italian]{varioref}
\usepackage[utf8]{inputenc}
\usepackage[nottoc]{tocbibind}
\usepackage{hhline}
\usepackage{diagbox}

\usepackage[affil-it]{authblk}
\usepackage{lineno,hyperref}
\modulolinenumbers[50]

\usepackage{dcolumn}
\usepackage{bm}
\usepackage[bf, font=small]{caption}
\usepackage{tabularx,ragged2e,booktabs}

\usepackage{algorithm}
\usepackage{algpseudocode}

\usepackage{mathrsfs}
\usepackage{forest,adjustbox}
\makeatletter
\def\BState{\State\hskip-\ALG@thistlm}
\makeatother

\usepackage[italian, english]{babel}
\usepackage{helvet}
\usepackage{mathtools} 

\usepackage{amssymb}
\usepackage{amsfonts}
\usepackage{numprint}
\usepackage{verbatim}
\usepackage{graphicx}
\usepackage{eucal}
\usepackage{rotating}
\usepackage[pdftex]{lscape}
\usepackage{longtable}

\usepackage{fancyhdr}
\usepackage{color}
\usepackage{bbm}
\usepackage{dsfont}
\makeatletter
\def\BState{\State\hskip-\ALG@thistlm}
\makeatother
\usepackage{tikz}
\usetikzlibrary{3d}
\usetikzlibrary{patterns}
\usetikzlibrary{calc}
\usetikzlibrary{arrows}

\usepackage{pgfplots}
\usepackage{pgfplotstable}
\usetikzlibrary{decorations.pathreplacing}
\makeatletter
\newcommand{\gettikzxy}[3]{%
	\tikz@scan@one@point\pgfutil@firstofone#1\relax
	\edef#2{\the\pgf@x}%
	\edef#3{\the\pgf@y}%
}
\makeatother

\usepackage{multirow}
\usepackage{etoolbox}
\newcommand{\T}[2][]{\boldsymbol{#1\mathcal{\MakeUppercase{#2}}}} 

\DeclareMathOperator*{\argmin}{arg\,min}

\usepackage[top=3.845cm, bottom=3.845cm, left=3.525cm, right=3.525cm]{geometry}

\usepackage{subcaption}

\begin{document}
	
		\title{A new multilayer network construction via Tensor learning}

	\author{Giuseppe Brandi\thanks{Corresponding author. Email: giuseppe.brandi@kcl.ac.uk} $^{,\dagger}$, T. Di Matteo$^{\dagger,\ddagger,\S}$}
	
	\affil{$\dagger$ Department of Mathematics, King's College London, The Strand, London, WC2R 2LS, UK\\
		$\ddagger$ Department of Computer Science, University College London, Gower Street, London, WC1E 6BT, UK\\
		$\S$ Complexity Science Hub Vienna, Josefstaedter Strasse 39, A 1080 Vienna, Austria}
	
	\providecommand{\keywords}[1]{\textbf{\textit{Keywords:}} #1}
		\date{}
	\maketitle              

		\begin{abstract}

		Multilayer networks proved to be suitable in extracting and providing dependency information of different complex systems. The construction of these networks is difficult and is mostly done with a static approach, neglecting time delayed interdependences. Tensors are objects that naturally represent multilayer networks and in this paper, we propose a new methodology based on Tucker tensor autoregression in order to build a multilayer network directly from data. This methodology captures within and between connections across layers and makes use of a filtering procedure to extract relevant information and improve visualization. We show the application of this methodology to different stationary fractionally differenced financial data. We argue that our result is useful to understand the dependencies across three different aspects of financial risk, namely market risk, liquidity risk, and volatility risk. Indeed, we show how the resulting visualization is a useful tool for risk managers depicting dependency asymmetries between different risk factors and accounting for delayed cross dependencies. The constructed multilayer network shows a strong interconnection between the volumes and prices layers across all the stocks considered while a lower number of interconnections between the uncertainty measures is identified.
			
		\end{abstract}


	\section{Introduction}
Network structures are present in different fields of research. Multilayer networks represent a widely used tool for representing financial interconnections, both in industry and academia \cite{musmeci2017multiplex} and has been shown that the complex structure of the financial system plays a crucial role in the risk assessment \cite{Musmeci2015risk,macchiati2019systemic}. A complex network is a collection of connected objects. These objects,  such as stocks, banks or institutions, are called nodes and the connections between the nodes are called edges, which represent their dependency structure. Multilayer networks extend the standard networks by assembling multiple networks ‘layers’ that are connected to each other via interlayer edges \cite{boccaletti2014structure} and can be naturally represented by tensors \cite{brandi2019unveil}. The interlayer edges form the dependency structure between different layers and in the context of this paper, across different risk factors. However, two issues arise:

\begin{itemize}\item[1] The construction of such networks is usually based on correlation matrices (or other symmetric dependence measures) calculated on financial asset returns. Unfortunately, such matrices being symmetric, hide possible asymmetries between stocks.
	
	\item[2] Multilayer networks are usually constructed via contemporaneous interconnections, neglecting the possible delayed cause-effect relationship between and within layers.
	
\end{itemize}

In this paper, we propose a method that relies on tensor autoregression which avoids these two issues. In particular, we use the tensor learning approach establish in \cite{Brandi_tensor} to estimate the tensor coefficients, which are the building blocks of the multilayer network of the intra and inter dependencies in the analyzed financial data. In particular, we tackle three different aspects of financial risk, i.e. market risk, liquidity risk, and future volatility risk. These three risk factors are represented by prices, volumes and two measures of expected future uncertainty, i.e. implied volatility at 10 days (IV10) and implied volatility at 30 days (IV30) of each stock. In order to have stationary data but retain the maximum amount of memory, we computed the fractional difference for each time series \cite{jensen2014fast}. To improve visualization and to extract relevant information, the resulting multilayer is then filtered independently in each dimension with the recently proposed Polya filter \cite{marcaccioli2019polya}. The analysis shows a strong interconnection between the volumes and prices layers across all the stocks considered while a lower number of interconnection between the volatility at different maturity is identified. Furthermore, a clear financial connection between risk factors can be recognized from the multilayer visualization and can be a useful tool for risk assessment. The paper is structured as follows. Section \ref{sec_ta} is devoted to the tensor autoregression. Section \ref{sec_er} shows the empirical application while Section \ref{sec_c} concludes. 

\section{Tensor regression} \label{sec_ta}

Tensor regression can be formulated in different ways: the tensor structure is only in the response or the regression variable or it can be on both. The literature related to the first specification is ample \cite{zhou2013,li2017} whilst the fully tensor variate regression received attention only recently from the statistics and machine learning communities employing different approaches \cite{Brandi_tensor,lock2017}. The tensor regression we are going to use is the Tucker tensor regression proposed in \cite{Brandi_tensor}. The model is formulated making use of the contracted product, the higher order counterpart of matrix product \cite{Brandi_tensor} and can be expressed as: 

\begin{equation}\label{tr}
\T{Y}=\T{A} + \langle \T{X},\T{B} \rangle_{(\T{I_x};\T{I_B})}+\T{E}
\end{equation}
\vspace{0.5pt}

where $\T{X}\in \mathbb{R}^{N \times  I_1 \times \cdots \times I_N}$ is the regressor tensor, $\T{Y} \in \mathbb{R}^{N \times J_1\times \cdots \times J_M} $ is the response tensor, $\T{E}\in \mathbb{R}^{N \times J_1\times\cdots \times J_M}$ is the error tensor, $\T{A}\in \mathbb{R}^{1 \times J_1\times\cdots \times J_M}$ is the intercept tensor while the slope coefficient tensor, which represents the multilayer network we are interested to learn, is $\T{B}\in \mathbb{R}^{I_1 \times \cdots \times I_N \times J_1 \times \cdots \times J_M}$. Subscripts $\T{I_x}$ and $\T{J_B}$ are the modes over winch the product is carried out.	In the context of this paper, $\T{X}$ is a lagged version of $\T{Y}$, hence $\T{B}$ represents the multilinear interactions that the variables in $\T{X}$ generate in $\T{Y}$. These interactions are generally asymmetric and take into account lagged dependencies being $\T{B}$ the mediator between two separate in time tensor datasets. Therefore, $\T{B}$ represents a perfect candidate to use for building a  multilayer network. However, the $\T{B}$ coefficient is high dimensional. In order to resolve the issue, a Tucker structure is imposed on $\T{B}$ such that it is possible to recover the original $\T{B}$ with smaller objects.\footnote{If the imposed Tucker rank is lower than the dimension of the tensor dataset, we have dimensionality reduction.} One of the advantages of the Tucker structure is, contrarily to other tensor decompositions such as the PARAFAC, that it can handle dimension asymmetric tensors since each factor matrix does not need to have the same number of components.
\subsection{Penalized Tensor regression}
Tensor regression is prone to over-fitting when intra-mode collinearity is present. In this case, a shrinkage estimator is necessary for a stable solution. In fact, the presence of collinearity between the variables of the dataset degrades the forecasting capabilities of the regression model. In this work, we use the Tikhonov regularization \cite{tikhonov1943stability}. Known also as Ridge regularization, it rewrites the standard Least Squares problem as	
\begin{equation}\label{optm2}
\T{\widehat{B}}=\argmin_{Trk(\T{B})\leq \T{R_{\bullet}}}\|\T{Y}- \langle \T{X},\T{B} \rangle_{(\T{I_x};\T{I_B})}\|^2_F + \lambda	\|\T{B}\|^2_F
\end{equation}
where $\lambda>0$ is the regularization parameter and $\| \|^2_F$ is the squared Frobenius norm. The greater the $\lambda$ the stronger is the shrinkage effect on the parameters. However, high values of $\lambda$ increase the bias of the tensor coefficient $\T{B}$. Indeed, the shrinkage parameter is usually set via data driven procedures rather than input by the user. The Tikhonov regularization can be computationally very expensive for big data problem. To solve this issue, \cite{arcucci2017decomposition} proposed a decomposition of the Tikhonov regularization. The learning of the model parameters is a nonlinear optimization problem that can be solved by iterative algorithms such as the Alternating Least Squares (ALS) introduced by \cite{kroonenberg1980principal} for the Tucker decomposition. This methodology solves the optimization problem by dividing it into small least squares problems. Recently, \cite{Brandi_tensor} developed an ALS algorithm for the estimation of the tensor regression parameters with Tucker structure in both the penalized and unpenalized settings. For the technical derivation refer to \cite{Brandi_tensor}.
\section{Empirical application: Multilayer network estimation}\label{sec_er}
In this section, we show the results of the construction of the multilayer network via the tensor regression proposed in Eq. \ref{tr}. 
\subsection{Data and fractional differentiation}\label{sec_data}
The dataset used in this paper is composed of stocks listed in the \textit{Dow Jones} (DJ). These stocks time series are recorded on a daily basis from 01/03/1994 up to 20/11/2019, i.e. 6712 trading days. We use 26 over the 30 listed stocks as they are the ones for which the entire time series is available. For the purpose of our analysis, we use log-differenciated prices, volumes, implied volatility at 10 days (IV10) and implied volatility at 30 days (IV30). In particular, we use the fractional difference algorithm of \cite{jensen2014fast} to balance stationarity and residual memory in the data. In fact, the original time series have the full amount of memory but they are non-stationary while integer log-differentiated data are stationary but have small residual memory due to the process of differentiation. In order to preserve the maximum amount of memory in the data, we use the fractional differentiation algorithm with different levels of fractional differentiation and then test for stationarity using the Augmented Dickey-Fuller test \cite{fuller2009introduction}. We find that all the data are stationary when the order of differentiation is $\alpha=0.2$. This means that only a small amount of memory is lost in the process of differentiation.

\subsection{Model selection}

The tensor regression presented in Eq. \ref{tr} has some parameters to be set, i.e. the Tucker rank and the shrinkage parameter $\lambda$ for the penalized estimation of Eq. \ref{optm2} as discussed in \cite{Brandi_tensor}. Regarding the Tucker rank, we used the full rank specification since we do not want to reduce the number of independent links. In fact, using a reduced rank would imply common factors to be mapped together, an undesirable feature for this application. Regarding the shrinkage parameter $\lambda$, we selected the value as follows. First, we split the data in a training set composed of $90\%$ of the sample and in a test set with the remaining $10\%$. We then estimated the regression coefficients for different values of $\lambda$ on the training set and then we computed the predicted $R^2$ on the test set. We used a grid of $\lambda=0,1,5,10,20,50.$ and the predicted $R^2$ is maximized at $\lambda=0$ (no shrinkage). 

\subsection{Results}	
\begin{figure}[h!]
	\begin{center}	
		\includegraphics[width=0.4955\textwidth,height=0.335\textheight]{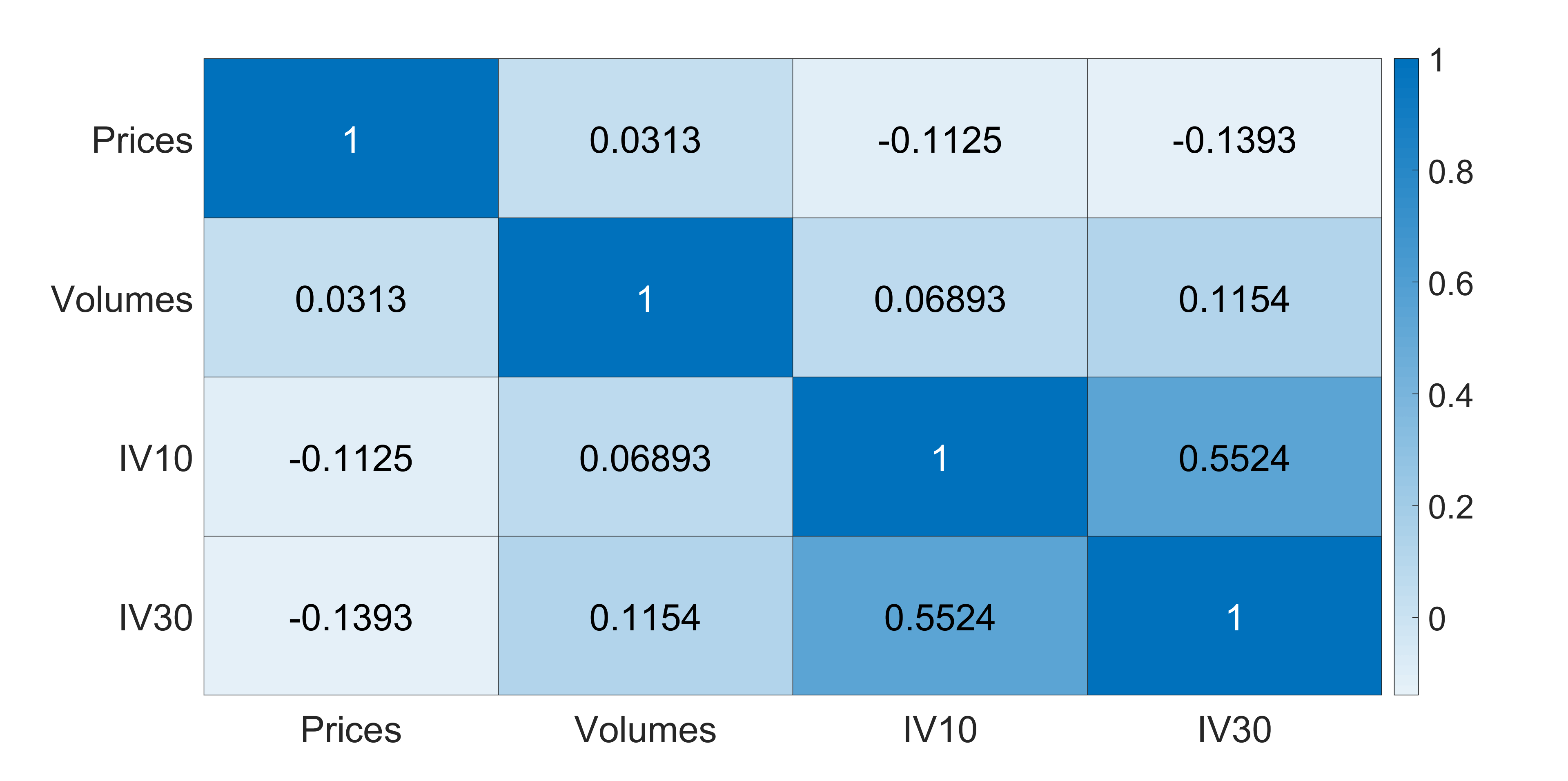} 
		\includegraphics[width=0.4955\textwidth,height=0.335\textheight]{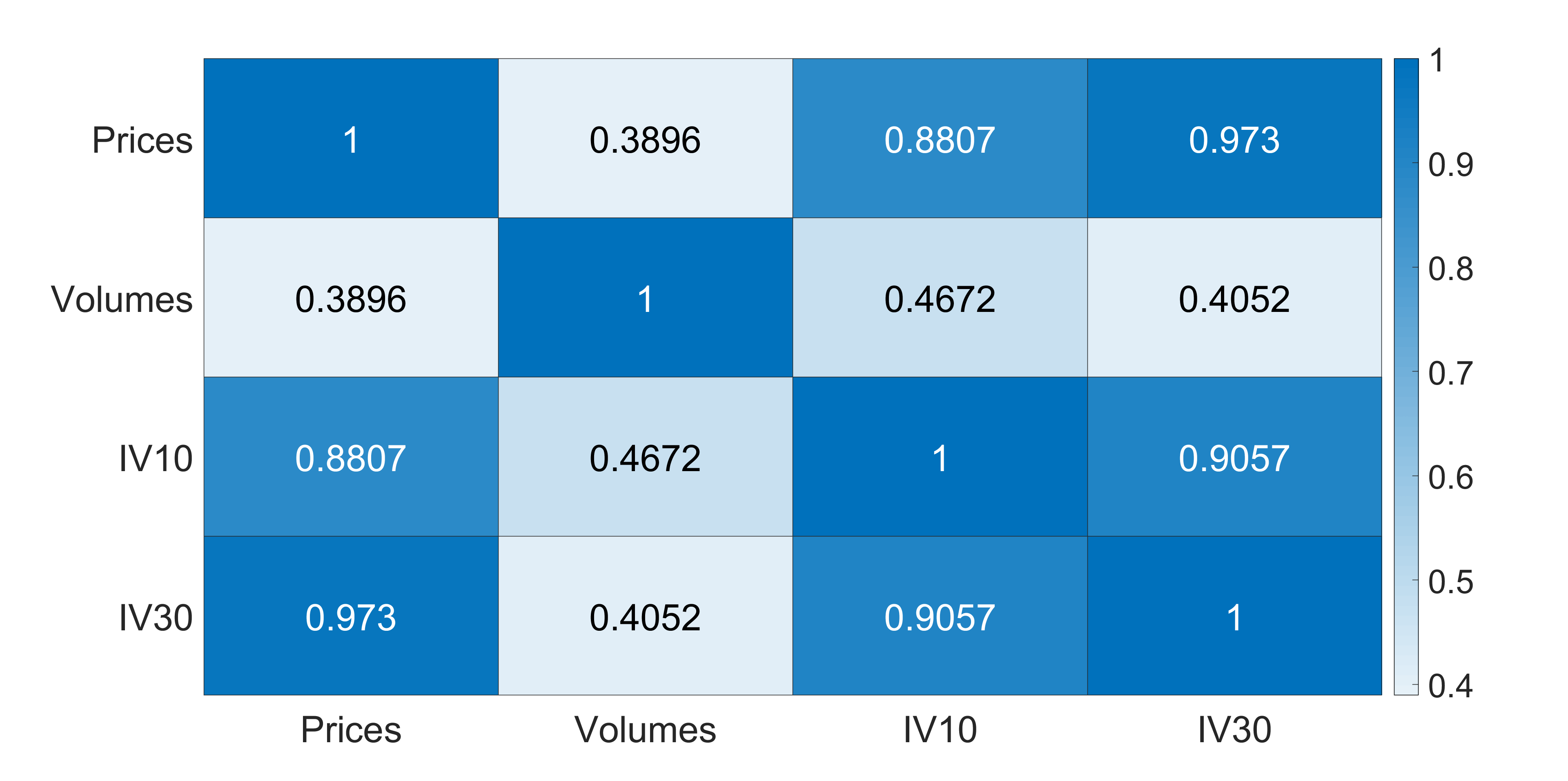} 
	\end{center}
	\caption{Multilayer network assortativity matrix and edge overlapping matrix. Linear scale.  Darker colour represents higher values.}
	\label{fig2}
	\vspace{5pt}
\end{figure}

In this section, we show the results of the analysis carried out with the data presented in Sec. \ref{sec_data}. The multilayer network built via the estimated tensor autoregression coefficient $\T{B}$ represents the interconnections between and within each layer. In particular $\T{B}_{i,j,k,l}$ is the connection between stock $i$ in layer $j$ and stock $k$ in layer $l$. It is important to notice that the estimated dependencies are in general not symmetric, i.e. $\T{B}_{i,j,k,l}\neq\T{B}_{k,j,i,l}$. However, the multilayer network constructed using  $\T{B}$ is fully connected. For this reason, a method for filtering those networks is necessary. Different methodologies are available for filtering information from complex networks\cite{marcaccioli2019polya,aste2005complex}. In this paper, we use the Polya filter of \cite{marcaccioli2019polya} as it can handle directed weighted networks and it is both flexible and statistically driven. In fact, it employs a tuning parameter $a$ that drives the strength of the filter and returns the p-values for the null null hypotheses of random interactions. We filter every network independently (both intra and inter connections) using a parametrization such that $90\%$ of the total links are removed.\footnote{Using hard thresholding the results are qualitatively equivalent.} In order to asses the dependency across the layers, we analyze two standard multilayer network measures, i.e. inter-layer assortativity and edge overlapping. A standard way to quantify inter-layer assortativity is to calculate Pearson’s correlation coefficient over degree sequences of two layers and it represents a measure of association between layers. High positive (negative) values of such measure mean that the two risk factors act in the same (opposite) direction. Instead, overlapping edges are the links between pair of stocks present contemporaneously in two layers. High values of such measure mean that the stocks have common connections behaviour. As it can be possible to see from Figure \ref{fig2}, prices and volatility have a huge portion of overlapping edges, still, these layers are disassortative as the correlation between the nodes sequence across the two layer is negative. This was an expected result since the negative relationship between prices and volatility is a stylized fact in finance. Not surprisingly, the two measures of volatility are highly assortative and have a huge fraction of overlapping edges.  

\vspace{5pt}

\begin{figure}[h!]
	\begin{center}	
		\includegraphics[height=0.6\textwidth, width=0.95\textwidth]{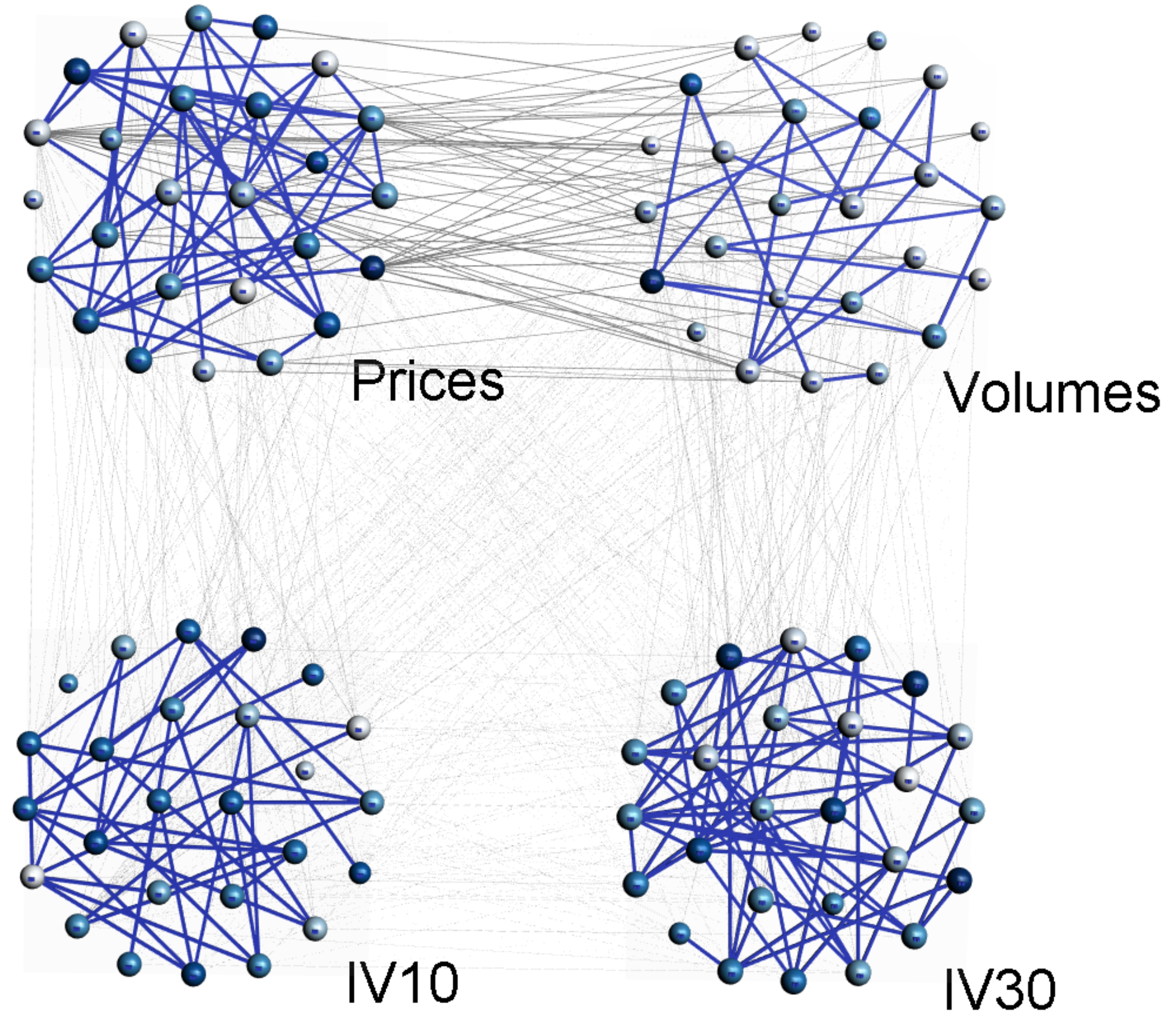}
		\caption{Estimated multilayer network. Node colours: loglog scale; darker colour is associated to higher strength of the node. Node size: loglog scale; darker colour is associated to higher k-coreness score. Edge colour: uniform.}
		\label{fig3}		
	\end{center}
\end{figure}

Finally, we show in Figure \ref{fig3} the filtered multilayer network constructed via the tensor coefficient $\T{B}$ estimated via the tensor autoregression of Eq. \ref{tr}. As it can be possible to notice, the volumes layer has more interlayer connections rather than intralayer connections. Since each link represents the effect that one variable has on itself and other variables in the future, this means that stocks' liquidity risk mostly influences future prices and expected uncertainty. The two volatility networks have a relatively small number of interlayer connections despite being assortative. This could be due to the fact that volatility risk tends to increase or decrease through a specific maturity rather than across maturities. It is also possible to notice that more central stocks, depicted as bigger nodes in Figure \ref{fig3}, have more connections but that this feature does not directly translate in a higher strength (depicted as darker colour of the nodes). This is a feature already emphasized in \cite{macchiati2019systemic} for financial networks. 

From a financial point of view, such graphical representation put together three different aspects of financial risk: market risk, liquidity risk (in terms of volumes exchanged) and forward looking uncertainty measures, which account for expected volatility risk. In fact, the stocks in the volumes layer are not strongly interconnected but produce a huge amount of risk propagation through prices and volatility. Understanding the dynamics of such multilayer network representation would be a useful tool for risk managers in order to understand risk balances and propose risk mitigation techniques.

\section{Conclusions}\label{sec_c}
In this paper, we proposed a methodology to build a  multilayer network via the estimated coefficient of the Tucker tensor autoregression of \cite{Brandi_tensor}. This methodology, in combination with a filtering technique, has proven able to reproduce interconnections between different financial risk factors. These interconnections can be easily mapped to real financial mechanisms and can be a useful tool for monitoring risk as the topology within and between layers can be strongly affected in distressed periods. In order to preserve the maximum memory information in the data but requiring stationarity, we made use of fractional differentiation and found out that the variables analyzed are stationary with differentiation of order $\alpha=0.2$. The model can be extended to a dynamic framework in order to analyze the dependency structures under different market conditions. 
	
	\bibliographystyle{unsrt}
	\bibliography{TAR_paper_new}
	\end{document}